\documentclass[PRL,twocolumn,showpacs,preprintnumbers,amsmath,amssymb]{revtex4}

\usepackage{graphicx}
\usepackage{dcolumn}
\usepackage{bm}

\begin{document}

\title{Suppression of Excitation and Spectral Broadening Induced by Interactions in a Cold Gas of Rydberg Atoms}

\author{Kilian Singer}%
 \email{kilian.singer@physik.uni-freiburg.de}
 \author{Markus Reetz-Lamour}
\author{Thomas Amthor}
\author{Luis Gustavo Marcassa}
\altaffiliation{Instituto de F\'isica de S\~ao Paulo, Caixa Postal
369, 13560-970, S\~ao Carlos, S\~ao Paulo, Brazil}
\author{Matthias Weidem\"uller}
\email{m.weidemueller@physik.uni-freiburg.de}

\affiliation{Physikalisches Institut der Universit\"at Freiburg,
Hermann-Herder-Str.3, 79104 Freiburg, Germany}
\homepage{http://quantendynamik.physik.uni-freiburg.de}

\date{\today}

\begin{abstract}
We report on the observation of ultralong range interactions in a
gas of cold Rubidium Rydberg atoms. The van-der-Waals interaction
between a pair of Rydberg atoms separated as far as 100,000 Bohr
radii features two important effects: Spectral broadening of the
resonance lines and suppression of excitation with increasing
density. The density dependence of these effects is investigated
in detail for the S- and P- Rydberg states with main quantum
numbers $\textsf{n}\sim60$ and $\textsf{n}\sim80$ excited by
narrow-band continuous-wave laser light. The density-dependent
suppression of excitation can be interpreted as the onset of an
interaction-induced local blockade.
\end{abstract}

\maketitle

With the advances in laser cooling and trapping, new perspectives
for the investigation of Rydberg atoms \cite{gallagher94} have
been opening. When cooled to very low temperatures, the core
motion can be neglected for the timescales of excitation (``frozen
Rydberg gas''). Unexpected effects have been discovered, such as
the many-body diffusion of excitation
\cite{mourachko98,stoneman87}, the population of
high-angular-momentum states through free charges \cite{dutta01},
or the spontaneous formation and recombination of ultracold
plasmas~\cite{robinson00,gallagher03}. Other fascinating features
of cold, interacting Rydberg atoms have been proposed but not been
observed so far, e.g. the creation of ultralong range
molecules~\cite{greene00,boisseau02}, whereas molecular crossover
resonances have already been found experimentally
\cite{farooqi03}. One outstanding property of Rydberg atoms is
their high polarizability, caused by the large distance between
the outer electron and the core. This leads to strong electric
field sensitivity and strong long-range dipole-dipole and
van-der-Waals (vdW) interactions are expected. First indications
of interaction effects between Rydberg gases at high densities
have been found in an atomic beam experiment \cite{raimond81} and,
more recently, collisional evidence for ultralong range
interactions in a cold Rydberg gas has been
reported~\cite{oliveira03}. In a frozen gas these interactions
make Rydberg atoms possible candidates for quantum information
processing~\cite{jaksch00,lukin01}. One promising approach is
based on the concept of a dipole blockade \cite{lukin01}, {\it
i.e.} the inhibition of multiple Rydberg excitations in a confined
volume due to interaction-induced energy shifts.

In this Letter we report on experimental evidence for ultralong
range interactions in a frozen Rydberg gas and we present
high-resolution spectroscopic signatures of these interactions.
Similar indications of suppressed excitation have recently been
observed in pulsed Rydberg excitation from a cold
gas~\cite{tong04}. Different to these findings, our experiment
makes use of a tunable narrow-bandwidth continuous-wave (cw) laser
for Rydberg excitation and thus allows for high-resolution
spectroscopy of the resonance lines. By varying the density of
Rydberg atoms in a controlled way, the influence of interactions
on the strength and the shape of these lines is investigated in
detail. Signatures of ultralong range interactions appear as
spectral broadening of the excitation lines and saturation of the
resonance peak height, the latter being an indication of the
dipole blockade.

To realize a frozen gas of Rydberg atoms we trap $^{87}$Rb atoms
in a dispenser-loaded magneto-optical trap (MOT).
The MOT is located inside a vacuum chamber (background pressure
$5\times10^{-11}\textrm{mbar}$) which provides optical access
along three main orthogonal axes and two diagonals. The MOT coils
are located inside the vacuum chamber. The trapping laser light is
provided by a diode laser system consisting of two diode lasers
which are simultaneously injection-locked to a
frequency-stabilized extended cavity diode laser (ECDL) at 780\,nm
and added coherently \cite{singer03}. The output of this
laser-addition setup is sent through a single-mode optical fiber
for mode cleaning providing about $70\,\textrm{mW}$ at the output.
The frequency is detuned by $1.5\,\Gamma$ (natural line width
$\Gamma/2\pi=6.1\;\textrm{MHz}$) to the red of the cooling and
trapping transition $\textrm{5S}_{1/2}(\textrm{F=2}) \rightarrow
\textrm{5P}_{3/2}(\textrm{F=3})$. A second ECDL drives the
repumping transition $\textrm{5S}_{1/2}(\textrm{F=1})\rightarrow
\textrm{5P}_{3/2}(\textrm{F=2})$. The MOT contains $1.2(3)\times
10^7$ atoms at temperatures around 100\,$\mu$K. The atom cloud can
be approximated by a Gaussian spatial distribution with an $1/e$
diameter of $1.2(0.3)\;\textrm{mm}$ resulting in peak densities of
$1.1(3)\times 10^{10}\;\textrm{cm}^{-3}$, as determined by
absorption imaging with a resonant probe laser beam~\cite{note1}.

\begin{figure}
\includegraphics[width=\columnwidth]{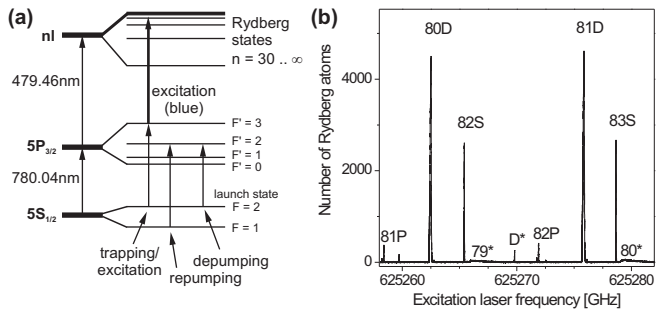}
\caption{\label{fig:setup}(a) Relevant energy levels of $^{87}$Rb
addressed by transitions driven by the trapping laser, the
repumping and the depumping lasers at 780\,nm, and the Rydberg
excitation laser at 479\,nm. (b) Rydberg spectrum at main quantum
number $\textsf{n}\sim80$. The small D* line is attributed to a
hyperfine ghost excited from the $\textrm{5S}_{1/2}(\textrm{F=1})$
state. The small structures labeled 79* and 80* are hydrogen-like
states with $\ell\geq3$.}
\end{figure}

The laser light for driving the transition to Rydberg states
$\textrm{5P}_{3/2}(\textrm{F=3})\rightarrow \textsf{n}\ell_{j}$ is
provided by a commercial laser system (Toptica, TA-SHG 110)
consisting of an ECDL at 960\,nm subsequently amplified to 1W and
then frequency doubled to 479\,nm with a line width
$<2\;\textrm{MHz}$. The output beam
which can be switched by an acousto-optical modulator (AOM) is
guided along one of the diagonal axes of the vacuum chamber and
focused to a waist of $80(10) \mu\textrm{m}$ at the center of the
atom cloud. The overlap between the atom cloud and the blue laser
beam defines an effective Rydberg excitation volume
$V_{\mathrm{exc}}$ = 1.1\,mm$^{3}$.

The two-step excitation into Rydberg states is schematically shown
in Fig.\,\ref{fig:setup}(a). At the beginning of an excitation
cycle, the trapping laser is tuned into resonance with the
$5S_{1/2}(F=2)\rightarrow 5P_{3/2}(F'=3)$ transition.
Subsequently, the blue Rydberg excitation laser which is tuned to
the
Rydberg manifold is switched on for typically $20\mu$s. Rydberg
atoms are accumulated in the excitation volume during this time,
since the lifetime of the Rydberg states exceeds 100\,$\mu$s
\cite{gallagher94,theodosiou00}. After the Rydberg excitation
laser is blocked, the trapping laser is reset to the MOT detuning
of $-1.5\,\Gamma$. The Rydberg atoms are field-ionized by applying
a voltage of 40 Volts to the central field plates consisting of
nickel meshes separated by 13.2\,mm with $95\%$ transparency
through which the MOT beams pass almost unperturbed. The ions are
accelerated onto a micro-channel plate detector (MCP). Residual
electrical fields during Rydberg excitation are mainly caused by
the MCP potential of $-1.9\,$kV. Field components perpendicular to
the field plates are compensated by applying a small voltage,
while parallel field components are measured to be less than
$0.16\;\textrm{V/cm}$. The excitation cycle is repeated every
$20\;\textrm{ms}$ while maintaining a continuously loaded
steady-state MOT. To take an excitation spectrum, the frequency of
the blue laser is scanned at a rate of 300 MHz/s.

The density of Rydberg atoms can be controlled by varying the
population in the $\textrm{5S}_{1/2}(F=2)$ launch state through
optical pumping into the $\textrm{5S}_{1/2}(F=1)$ state.
For this purpose, the repumping laser is attenuated by an AOM
while an additional depumping laser resonant with the transition
$\textrm{5S}_{1/2}(\textrm{F=2})\rightarrow
\textrm{5P}_{3/2}(\textrm{F=2})$ (see Fig.~\ref{fig:setup}(a)) is
switched on. Within $1\,$ms, the density of atoms in the launch
state is lowered by one order of magnitude, as is measured by
recording the fluorescence on the closed $\textrm{5S}_{1/2}(F=2)$
- $\textrm{5P}_{3/2}(F=3)$ transition with a photodiode. Note that
no atoms are lost from the MOT capture volume during this short
period of time, and that the excitation volume $V_{\mathrm{exc}}$
remains unaltered. By delaying the two-step Rydberg excitation
scheme with respect to the start of the depumping process, one can
thus modify the density of atoms in the launch state without
changing the total density of atoms in the MOT.

Absolute numbers of Rydberg atoms per excitation cycle are
obtained by comparing the number of atoms lost from the trap, as
deduced from the change in fluorescence rate at 780\,nm, with the
integral of the ion signal at the MCP. In this way, the response
of the MCP can be calibrated and nonlinearities at large ion
signals are compensated. Throughout this paper, the number of
Rydberg atoms is derived from the number of measured ions at the
MCP after calibration. Peak Rydberg densities
$\hat{n}_\mathrm{Ryd}$ are calculated by dividing the number of
Rydberg atoms by the effective excitation volume
$V_{\mathrm{exc}}$, assuming that the spatial distribution of
Rydberg atoms follows the distribution of ground state atoms.
We estimate the systematic error for the absolute number and peak
density of Rydberg atoms to be a factor of roughly $3$.

A Rydberg spectrum for $\textsf{n}\sim80$ is depicted in
Fig.~\ref{fig:setup}(b). 
About 4500 Rydberg atoms are detected for the strong
dipole-allowed D transitions. The resonance line width is
typically $\sim30\;\textrm{MHz}$ caused by saturation broadening
of the first excitation step~\cite{note2}. Dipole-forbidden P
states are also excited due to the residual electrical field.
Under our experimental conditions, the ion formation rate for
$\textsf{n}\sim80$ states through blackbody radiation
\cite{gallagher94} is estimated from calculated lifetimes
\cite{theodosiou00} to be approximately $200\textrm{Hz}$.
Ionization rates due to collisions of cold Rydberg atoms with
residual hot Rydberg atoms are of the same order of magnitude
assuming an upper limit of $10^{8}\;\textrm{cm}^{-3}$ for the
densities of hot Rydberg atoms. We therefore do not expect
significant effects of spontaneous creation of ions and avalanche
processes during the timescales of excitation ($\sim20\;\mu$s)
\cite{robinson00}.

\begin{figure}
\includegraphics[width=\columnwidth]{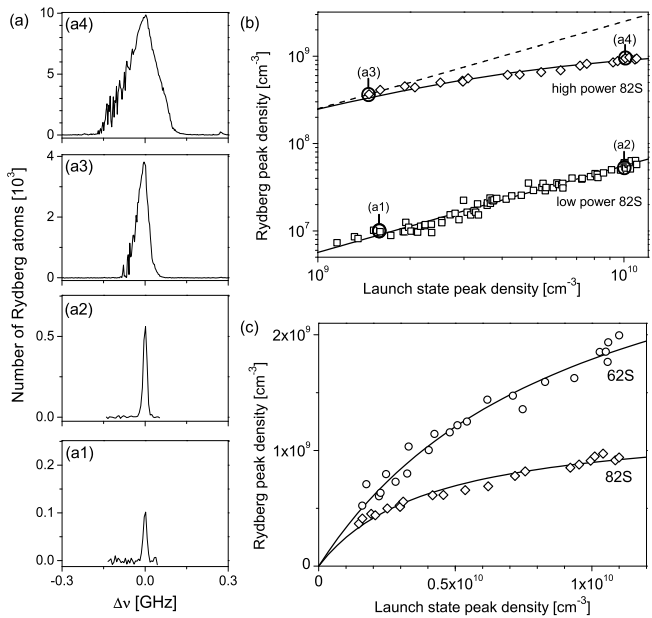}
\caption{\label{fig:blockade}(a) Excitation spectra of the
$\textrm{82S}_{1/2}$ state at different intensities of the blue
excitation laser [$6\,\textrm{W/cm}^2$ for (a1) and (a2),
$500\;\textrm{W/cm}^2$ for (a3) and (a4)] and launch state
densities [(a1) $1.6\times10^9\;\textrm{cm}^{-3}$, (a2)
$1.0\times10^{10}\;\textrm{cm}^{-3}$, (a3)
$1.5\times10^9\;\textrm{cm}^{-3}$, (a4)
$1.0\times10^{10}\;\textrm{cm}^{-3}$]. (b) Rydberg peak densities
on the $\textrm{82S}_{1/2}$ resonance versus launch state density
for $6\;\textrm{W/cm}^2$ ($\Box$) and $500\;\textrm{W/cm}^2$
($\diamondsuit$). The dashed line is a linear extrapolation from
the origin through the first data point. Data points corresponding
to the spectra (a1) to (a4) are marked. (c) Comparison of the
density-dependent suppression of Rydberg excitation for the
$\textrm{82S}_{1/2}$ (same data as in (b)) and
$\textrm{62S}_{1/2}$ ($\bigcirc$) resonances. The
$\textrm{62S}_{1/2}$ data was taken at $350\;\textrm{W/cm}^2$. The
solid lines in (b) and (c) show fitted saturation functions (see
text).}
\end{figure}

To study interaction effects between Rydberg atoms, excitation
spectra of the $\textrm{82S}_{1/2}$ state are recorded for
different Rydberg densities. The density is changed by either
varying the power of the excitation laser or, as described above,
by changing the density $\hat{n}_\mathrm{g}$ of atoms in the
$5\textrm{S}_{1/2}(F=2)$ launch state.
Figs.~\ref{fig:blockade}(a1) and \ref{fig:blockade}(a2) are taken
at low intensity of the blue excitation laser ($I=
6\;\textrm{W/cm}^2$). 
In this ``low power regime'', the line shows no broadening.
The number, and thus the density, of Rydberg atoms increases with
$\hat{n}_\mathrm{g}$, as the number of launch state atoms in the
fixed excitation volume increases.
The behavior changes drastically at higher intensity of the blue
excitation laser. 
Figs.~\ref{fig:blockade}(a3) and (a4) show the resonance line at
an excitation intensity of about $500\;\textrm{W/cm}^2$.
In the ``high power'' regime characterized by Rydberg densities in
the $10^8\;\textrm{cm}^{-3}$ range, one observes an asymmetric
broadening of the line. Qualitatively similar line broadenings at
Rydberg densities around $10^{11}\;\textrm{cm}^{-3}$ were observed
earlier in an atomic beam experiment~\cite{raimond81}. The line
broadenings are attributed to level shifts of Rydberg states
induced by the long-range vdW interactions of Rydberg atom pairs,
although an appropriate model describing these line shapes is
still lacking. Our high-resolution spectra in a cold gas reveal
additional spectral features in the red wing of the resonance
line, which show strong shot-to-shot fluctuations. Since these
features never appear at the blue side of the resonance,
independent of the scanning direction, one is tempted to attribute
them to molecular resonances in the attractive part of an
interaction potential. The origin of the fluctuating resonances
remains to be clarified.


The line broadening is accompanied by suppression of excitation on
resonance. In Fig.~\ref{fig:blockade}(b) we have plotted the peak
density of Rydberg atoms excited on resonance versus peak density
$\hat{n}_\mathrm{g}$ of $5\textrm{S}_{1/2}(F=2)$ atoms for both
the ``low power'' and the ``high power'' regimes. If no
interaction was present, the peak density of Rydberg atoms scales
linearly with $\hat{n}_\mathrm{g}$. This is indeed realized in the
``low power'' regime as a fit to the data yields
$\hat{n}_\mathrm{Ryd} = p_{\mathrm{82,low}} \,
\hat{n}_\mathrm{g}^{0.99(2)}$ with a probability
$p_{\mathrm{82,low}} = 0.006(1)$ for an atom to become excited in
a Rydberg state during the 20-$\mu$s excitation interval. In the
``high power'' regime, however, the increase of Rydberg density
scales less than linear with $\hat{n}_\mathrm{g}$ clearly showing
a suppression of excitation. At high excitation rate and density,
we detect only $10^4$ Rydberg atoms,
which is a factor of $\sim2.7$ less than expected from simple
linear density scaling (see dashed line in
Fig.~\ref{fig:blockade}(b)). The data is fitted by a heuristic
saturation function of the form $\hat{n}_\mathrm{Ryd} =
p_{\mathrm{82,high}} \, \hat{n}_\mathrm{g} /
(1+\frac{\hat{n}_\mathrm{g}}{n_\mathrm{sat}})$ giving
$p_{\mathrm{82,high}}=0.31(2)$ and
$n_\mathrm{sat}=4.1(4)\times10^9$\,cm$^{-3}$. The deviation of
$p_{\mathrm{82,high}}/p_{82,\mathrm{low}} = 56$ from the ratio of
the blue excitation laser power of 83 indicates a slight power
saturation of the Rydberg excitation. It is important to note that
power saturation can neither explain the asymmetric broadening of
the excitation lines nor can it explain the
\emph{density-dependent} saturation of the Rydberg excitation.
Saturation caused by the MCP detection is excluded by the
calibration and linearization procedure. Therefore we attribute
the suppression of the Rydberg excitation to interatomic
interactions resulting in the onset of a dipole blockade caused by
vdW interactions.

To further test this interpretation we have compared the
saturation for different main quantum numbers $\textsf{n}$. The
vdW interaction potential between two Rydberg atoms strongly
increases with $\textsf{n}$. Therefore, one expects weaker
suppression of Rydberg excitation for lines with lower main
quantum number, although the transition strength of these lines is
much higher as it scales with $(\textsf{n}^*)^{-3}$ where
$\textsf{n}^*$ denotes the main quantum number corrected by the
corresponding quantum defect \cite{gallagher94}.
Fig.~\ref{fig:blockade}(c) shows a comparison of resonantly
excited Rydberg atoms on the 82S$_{1/2}$ line
and the 62S$_{1/2}$ line. The solid lines show fits of the
saturation function described above. For the 62S$_{1/2}$ line we
find $p_{\mathrm{62,high}} = 0.37(2)$ and $n_{\textrm{sat}} =
9(1)\times10^9\;\textrm{cm}^{-3}$.
The 62S$_{1/2}$ saturation density is larger by more than a factor
of 2 than the saturation density of the 82S$_{1/2}$ excitation.
Additionally, the asymptotic Rydberg density ($p_{\textrm{ryd}}
n_{\textrm{sat}}$ for $\hat{n}_g\gg n_{\textrm{sat}})$) for
62S$_{1/2}$ relative to 82S$_{1/2}$  is larger by factor of $2.8$.
Both findings support the interpretation in terms of an
interaction-induced blockade.

\begin{figure}
\includegraphics[width=\columnwidth]{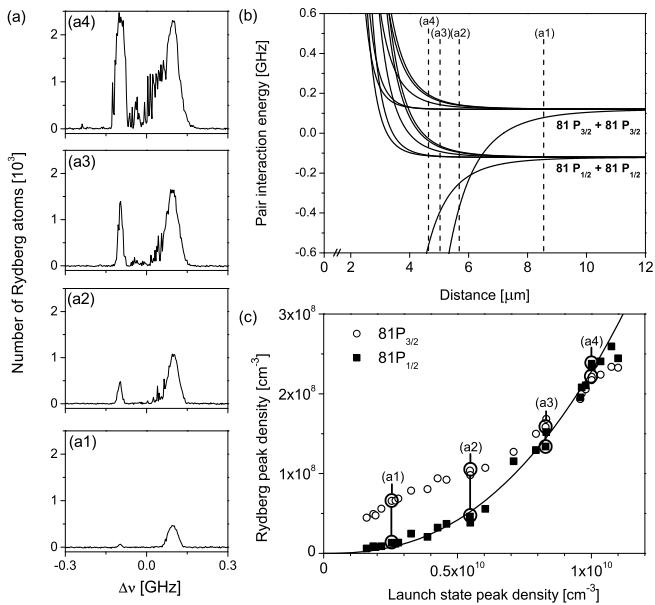}
\caption{\label{fig:pscaling}(a) Resonance line for the fine
structure doublet $\textrm{81P}_{1/2}$ (left) and
$\textrm{81P}_{3/2}$ (right) at excitation intensity of about
$500\;\textrm{W/cm}^2$ and different launch state densities [(a1)
$2.5\times10^9\;\textrm{cm}^{-3}$, (a2)
$5.5\times10^{10}\;\textrm{cm}^{-3}$, (a3)
$8.3\times10^9\;\textrm{cm}^{-3}$, (a4)
$1.0\times10^{10}\;\textrm{cm}^{-3}$]. (b) Calculated vdW pair
potentials for the $\textrm{81P}_{1/2}+\textrm{81P}_{1/2}$ and
$\textrm{81P}_{3/2}+\textrm{81P}_{3/2}$ asymptote versus the
interatomic distance. Dashed lines indicate the average nearest
neighbor distances $\hat{n}_{\mathrm{g}}^{-1/3}$ for the
corresponding spectra in (a). (c) Resonance peak densities of the
$\textrm{81P}_{1/2}$ and the $\textrm{81P}_{3/2}$ line as a
function of the launch state peak density. The solid line is a
polynomial fit to the data. The data points corresponding to the
graphs depicted in (a) are marked. }
\end{figure}

We have also studied the density dependence of the
dipole-forbidden fine structure doublet $\textrm{81P}_{1/2}$ and
$\textrm{81P}_{3/2}$ (left and right peak in
Fig.~\ref{fig:pscaling}(a), respectively), for which theoretical
calculations of the pair potentials are available
~\cite{robin04,marinescu97}.
Two features are remarkable: With increasing density the
$\textrm{81P}_{3/2}$ peak develops a wing on the red side of the
resonance, again with the fluctuating resonances (see also
Fig.~\ref{fig:blockade}(a)), while the $\textrm{81P}_{1/2}$ peak
is hardly broadened, but grows much faster with the launch state
density.
As shown in Fig.~\ref{fig:pscaling}(b), the attractive
$81\textrm{P}_{3/2}$ interaction asymptote is much steeper than
the $81\textrm{P}_{1/2}$ asymptote due to a coincidental
near-resonant enhancement in the $C_6$ coefficient. This explains
why broadening is only observed for the red wing of the
$\textrm{81P}_{3/2}$ line but not for the $\textrm{81P}_{1/2}$
line. The average nearest neighbor distances derived from the
launch state density are depicted by the dashed lines in
Fig.~\ref{fig:pscaling}(b). The red broadening starts when the
average nearest neighbor distance becomes comparable to
interatomic distances at which the attractive interaction
potential starts to bend, and observed broadening and interaction
energy are within the same order of magnitude.

Contrary to the dipole-allowed S and D lines, the dipole-forbidden
$\textrm{81P}_{1/2}$ and $\textrm{81P}_{3/2}$ peaks grow much
faster than linear with the launch state density
$\hat{n}_{\mathrm{g}}$, as shown in Fig.~\ref{fig:pscaling}(c). A
fit to the data of the $\textrm{81P}_{1/2}$ resonance yields a
scaling with $\hat{n}_{\mathrm{g}}^{2.4(1)}$. The behavior of the
$\textrm{81P}_{3/2}$ peak is more complicated, but shows
saturation relative to the $\textrm{81P}_{1/2}$ peak, which may
again be caused by a interaction-induced blockade effect. Whether
the density dependence of the dipole-forbidden P lines may be
attributed to interaction-induced S- and D-admixtures in addition
to the admixture caused by the residual static electric field
($<0.16$\,V/cm) has to be clarified in the framework of a detailed
model for the cw Rydberg excitation in a cold gas.

In conclusion, we have explored two signatures of interactions in
a cold Rydberg gas by controlling the density: A broadening of
Rydberg resonance lines with additional spectral features in the
red wings and a suppression of on-resonance excitation. We find
qualitative agreement between the observed broadening with
predicted pair-interaction vdW potentials in the case of the 81P
fine-structure doublet. Theoretical calculations of vdW potentials
including fine structure are needed to allow for a more
quantitative description of the observed line shapes.


The saturation of Rydberg excitation with increasing launch state
density is a complementary signature of the ultralong range
Rydberg-Rydberg interactions and marks the onset of a vdW-induced
dipole blockade. While the spectral broadening of the resonance
lines can only be understood as an ``instantaneous'' excitation of
Rydberg pairs, it is not clear whether the density-dependent
saturation occurs in a similar way or whether it is induced by a
process based on the successive creation of Rydberg atoms. The
latter hypothesis has successfully been employed in the
interpretation of similar saturation effects recently observed for
Rydberg atoms excited through short duration, high intensity laser
pulses from a cold gas \cite{tong04}. Our experimental approach
employing narrow-band cw Rydberg excitation can directly be
applied to the implementation of quantum information processing
with Rydberg atoms~\cite{lukin01,Teo03} by further increasing the
density, thus proceeding from the two-body limit to the full
many-body regime of the dipole blockade.

The project is supported in part by the Landes\-stiftung
Baden-W\"urttemberg in the framework of the ''Quantum Information
Processing'' program and by the DAAD-PROBRAL program. We
acknowledge contributions from S.~F\"olling and M.~Tscherneck. We
are also indebted to D.~Schwalm for generous support at the
Max-Planck-Institute for Nuclear Physics. We thank R.~C\^ot\'e,
P.~Gould and E.~Eyler for valuable discussions and providing us
with the 81P potentials.

\bibliography{rydberginteractionetal}

\begin{thebibliography}{21}
\bibitem[{not({\natexlab{a}})}]{note1}
\bibinfo{note}{We estimate a factor of 2 as the systematic error for the
  determination of the density and number of atoms.}
\bibitem[{not({\natexlab{b}})}]{note2}
\bibinfo{note}{Broadening by the Zeeman shift in the inhomogeneous MOT magnetic
  field is below 5\,MHz.}
\end{thebibliography}

\end{document}